\pdfoutput=1

\documentclass[pra,10pt,showpacs,superscriptaddress,footnoteinbib]{revtex4}
\usepackage{amsmath}
\usepackage{latexsym}
\usepackage{amssymb}
\usepackage{graphics,epstopdf}
\usepackage[colorlinks=true, citecolor=blue, urlcolor=blue ]{hyperref}
\usepackage{epsf,graphics,graphicx}

\newcommand{\n}{\noindent}

\newcommand{\ed}{\end{document}}
\newcommand{\beq}{\begin{equation}}
\newcommand{\eeq}{\end{equation}}

\begin{document}

\title{Inertial effect on spin orbit coupling and spin transport}

\author{B. Basu\footnote{Electronic
address: {sribbasu@gmail.com}}${}^{}$ and Debashree Chowdhury\footnote{Electronic address:{debashreephys@gmail.com}}${}^{}$} \affiliation{Physics and
Applied Mathematics Unit, Indian Statistical Institute, 203
B.T.Road, Kolkata 700 108, India}


\begin{abstract}
\n
We theoretically study the renormalization of inertial effects on the spin dependent transport of conduction electrons in a semiconductor by taking into account the interband mixing on the basis of $\vec{k}.\vec{p}$ perturbation theory. In our analysis, for the generation of spin current  we have used the extended Drude model where the spin orbit coupling plays an important role. We predict enhancement of the spin current resulting from the rerormalized spin orbit coupling  effective in our model in cubic and non cubic crystal.  Attention has been paid to clarify  the importance of  gauge fields in the spin transport of this inertial system. A theoretical proposition of a perfect spin filter has been done through the  Aharonov-Casher like phase corresponding to this inertial system. For a time dependent acceleration, effect of  $\vec{k} . \vec{p}$ perturbation on the spin current and spin polarization has also been addressed. Furthermore, achievement of a tunable source of polarized spin current through  the non uniformity of the inertial spin orbit coupling strength has also been discussed.

\end{abstract}


\maketitle

\section{Introduction}
In semiconductor band structure spin-orbit coupling (SOC), which originates from the relativistic coupling of spin and orbital motion of electrons, plays a very important role from the perspective of spin Hall effect. Understanding the effect of SOI is indispensable in the study of $spin$ $current$, a flow of spins.
Although spin current can be induced easily, detection and control of spin current is a challenging research area both for theoretical and experimental physicists and has attracted a lot of attention in the field of spintronics \cite{wolf,zutic,sh1}.
 Spintronics aims to use the spin properties of electrons along with the charge degrees of freedom and
 has emerged as the most pursued area in condensed matter physics and nanotechnology. In this regard,
 the theoretical prediction of the spin Hall effect (SHE) \cite{spinh} and its application to spintronics has seen
 considerable advancement. This effect is observed experimentally in semiconductors \cite{rashba, winkler}
and metals\cite{sh2}.

Though studies on the inertial effect of electrons has a long standing history \cite{barnett,ein,tol,o} but the contribution of the spin-orbit interaction (SOI) in accelerating frames has not much been addressed in the literature \cite{cb,cbs}. Recently, a theory has been proposed \cite{matsuo new,c} describing the direct coupling of mechanical rotation and spin  where the generation of spin current arising from rotational motion has been predicted.
Inclusion of the inertial effects in semiconductors can open up some fascinating phenomena, yet not addressed.
So it is appealing to investigate how the inertial effect affects some aspects of spin transport in semiconductors.
In addition, the role of SOI in  connection to spin Hall effect may inspire one to study
the gauge theory of this inertial spin orbit Hamiltonian.

The spin dynamics of the semiconductor is influenced by the $\vec{ k} . \vec{ p}$ perturbation theory as the band structure of a semiconductor in the vicinity of the band edges can be very well described by the $ \vec{ k} . \vec{ p}$ method.
On the basis of $ \vec{ k} . \vec{ p}$ perturbation theory, by taking into account the interband mixing, one can reveal many characteristic features related to spin dynamics.
In this paper, we theoretically investigate the generation of spin current in a solid on the basis of
$\vec{ k} . \vec{ p}$ perturbation \cite{winkler} with a generalized spin orbit Hamiltonian which includes the inertial effect due to acceleration. The generation of spin current is studied in the extended Drude model framework, where the spin orbit coupling has played an important role.  It is shown in our present paper that spin current appearing  due to the combined action of the external electric field, crystal field  and the induced  inertial electric field via the total effective spin-orbit interaction is enhanced by the interband mixing of the conduction and valence band states. We have also studied the Aharonov-Casher like phase which corresponds to the effective SOI present in the model. Through the interplay of Aharonov-Bohm phase ($AB$) and Aharonov-Casher ($AC$) phases, we are able to propose a perfect spin filter for the accelerating system. Also by taking into consideration of a special profile of the acceleration in a trilayer system, we can set up a tunable spin filter. Renormalization of the spin current and spin polarization for the time dependent acceleration has also been investigated.
Here we consider the $\vec{ k} . \vec{p}$ perturbation in the $8 \times 8$ Kane model and write the total Hamiltonian including the inertial effect due to acceleration.

The paper is organized as follows. In Section II we write the total Hamiltonian of the $8\times 8$ Kane model with
$\vec{k} . \vec{p}$ perturbation including the effect of acceleration. Section III deals with the generation of the spin
current and conductivity in the semiconductors with cubic and non-cubic symmetry. The effect of time dependent acceleration on spin current and conductivity is discussed in section IV. The details on the gauge theory of our model, particularly the $AC$ phase, perfect spin filter and tunable spin filter is narrated in section V. Finally we conclude with section VI.

\section{Inertial spin orbit Hamiltonian and $\vec{k}.\vec{p}$ method}
We start with the Dirac Hamiltonian for a particle with charge $e$ and mass $m$ in an arbitrary non-inertial frame with constant linear acceleration $\vec{a}$  and without rotation which is given by \cite{o},
\begin{equation}\label{hamil00}
H_{I} = \beta mc^{2} +  c\left({\bf\alpha}.(\vec{p}-\frac{e\vec{A}}{c})\right)\\
+\frac{1}{2c}\left[(\vec{a}.\vec{r})((\vec{p}-\frac{e\vec{A}}{c}).\vec{\alpha})
((\vec{p}-\frac{e\vec{A}}{c}).\vec{\alpha})(\vec{a}.\vec{r})\right]
+\beta m
(\vec{a}.\vec{r}) + eV(\vec{r}),
\end{equation}
where the subscript $I$ in Hamiltonian (\ref{hamil00}) is due to the effect of inertia.
 Applying  a series of Foldy-Wouthuysen (FW) transformations \cite{gre,m} on the Hamiltonian(\ref{hamil00} ) we can write the Pauli-Schrodinger
 Hamiltonian for the two component electron wave function
 in the low energy limit as 
\begin{equation}
H_{FW} = \left( mc^{2} + \frac{(\vec{p}-\frac{e\vec{A}}{c})^{2}}{2m}\right) + eV(\vec{r}) +  m (\vec{a}.\vec{r})
-\frac{e\hbar}{2mc}\vec{\sigma}.\vec{B}\\-
\frac{e\hbar}{4m^{2}c^{2}}\vec{\sigma}.(\vec{E}\times \vec{p})
+\frac{\beta\hbar}{4mc^{2}}\vec{\sigma}.(\vec{a}\times \vec{p})\label{w1}
\end{equation}

where $\vec{E}$ and $\vec{B}$ are the external electric and magnetic field respectively. 

In the right hand side of Hamiltonian (\ref{w1}), the third term is an inertial potential term arising due to the acceleration $\vec{a}$. This potential $\displaystyle{V_{\vec{a}}(\vec{r}) = -\frac{m}{e}\vec{a}.\vec{r}}$ induces an electric field  $\displaystyle{\vec{E}_{\vec{a}} = \frac{m}{e}\vec{a}}$ \cite{c,cb}.
The induced electric field $\vec{E}_{\vec{a}}$ produces an inertial SOI term (sixth term in the right hand side of (\ref{w1})) apart from the SOI term due to the external electric field (fifth term in the right hand side of (\ref{w1})). The Hamiltonian (\ref{w1}) thus can be rewritten in terms of $\vec{E}_{\vec{a}}$ and $V_{\vec{a}}(\vec{r})$ as
\begin{equation}\label{hfw2}
H_{FW} = \left(mc^{2} + \frac{(\vec{p}-\frac{e\vec{A}}{c})^{2}}{2m}\right)\\ + e(V(\vec{r})- V_{a}(\vec{r}))-\frac{e\hbar}{2mc}\vec{\sigma}.\vec{B}
-
 \frac{e\hbar}{4m^{2}c^{2}}\vec{\sigma}.((\vec{E}-E_{\vec{a}})\times \vec{p}).
\end{equation}
In the above calculations we have neglected the $\frac{1}{c^{4}}$ terms and the
terms due to red shift effect of kinetic energy. The generalized spin-orbit interaction
$ \displaystyle{ \frac{e\hbar}{4m^{2}c^{2}}\vec{\sigma}.\left((\vec{E} -\vec{E}_{\vec a})\times \vec{p}\right)}$
 effective in our inertial system plays a significant  role in our analysis.

 We are interested in an effective Hamiltonian describing the motion of electrons in a solid incorporating the inertial effect due to acceleration. It is known that the physical parameters present in any Hamiltonian in vacuum are renormalized when considered in a solid. In an inertial frame, such renormalization effects in a crystalline solid is generally studied in the framework of $\vec{k} . \vec{p}$ perturbation theory using the Bloch eigenstates. One can renormalize the effect of acceleration on the basis of $\vec{k} . \vec{p}$ perturbation and the
 $8 \times 8$ Kane model \cite{kane}.

The basic idea of the Kane model is that the band edge eigenstates constitute a complete basis and to obtain the eigenstates away from the band edge the wave function is expanded in the band edge states, which gives rise to an $8 \times 8$
band Hamiltonian. Bands that are far away in energy can be neglected.
In presence of magnetic field the crystal momentum is given by $ \hbar\vec{k} = \vec{p} - q \vec{A}. $
The $\vec{ k} . \vec{ p}$ method leads to high-dimensional Hamiltonians, for example, an $8 \times 8$ matrix for the Kane model \cite{winkler}.

To this end, we start with a Hamiltonian of the well known $ 8\times 8 $ Kane model which takes into account the $\vec{ k} . \vec{ p}$ coupling between the $\Gamma_{6} $ conduction band and $\Gamma_{8} $ and $\Gamma_{7} $ valance bands which is given by
\begin{eqnarray}
H_{8 \times 8}  =  \left( \begin{array}{ccr}
H_{6c6c} & H_{6c8v} & H_{6c7v} \\
H_{8v6c} & H_{8v8v} & H_{8v7v}\\
H_{7v6c} & H_{7v8v} & H_{7v7v}
\end{array} \right)~~~~ \\
~~=  \left( \begin{array}{ccr}
(E_{c} + eV_{tot})I_2 & \sqrt{3}P\vec{ T} . \vec{ k} & -\frac{P}{\sqrt{3}}\vec{ \sigma} . \vec{ k} \\
\sqrt{3}P \vec{ T}^{\dag} . \vec{ k} & (E_{v} + eV_{tot})I_4  & 0 \\
-\frac{P}{\sqrt{3}}\vec{ \sigma} . \vec{ k} & 0  & (E_{v} - \triangle_{0} + eV_{tot})I_{2}
\end{array} \right)
\end{eqnarray}
Here, $V_{tot} = V(\vec{r})-V_{a}(\vec{r}),$ $E_{c}$ and $E_{v}$ are the energies at the conduction and valence band edges respectively. $ \triangle_{0}$ is the spin orbit gap, P is the Kane momentum matrix element which couples $s$ like conduction bands with $p$ like valence bands. This Kane Momentum matrix is almost constant for group III to V semiconductors, whereas $ \triangle_{0}$ and $E_{G}=E_c-E_v$ varies with materials. The $\vec{T}$ matrices are given as
\begin{equation}
T_{x}  = \frac{1}{3\sqrt{2}} \left( \begin{array}{ccrr}
-\sqrt{3} & 0 & 1 & 0 \\
0 & -1 & 0 & \sqrt{3}
\end{array} \right),~~~~
T_{y}  = -\frac{i}{3\sqrt{2}} \left( \begin{array}{ccrr}
\sqrt{3} & 0 & 1 & 0 \\
0 & 1 & 0 & \sqrt{3}
\end{array} \right),
T_{z}  = \frac{\sqrt{2}}{3} \left( \begin{array}{ccrr}
0 & 1 & 0 & 0 \\
0 & 0 & 1 & 0
 \end{array}\right)
\end{equation}
and $I_{2}, I_{4}$ are unit matrices of size $2$
and $4$ respectively.

It may be noted here that as the effect of rotation \cite{c} is not considered, the crystal momentum used in $\vec{k} . \vec{p}$ perturbation is not modified in our model. The effect of acceleration changes the electric potential $V(\vec{r})$ as well as the electric field $\vec{E}$. In our framework, the total potential and total electric field have been modified as  $V_{tot}(\vec{r})$ and  $\vec{E}_{tot}$ respectively.

The Hamiltonian (5) can now be reduced  to an effective Hamiltonian of the conduction band electron states \cite{winkler} in presence of acceleration as
\beq
H_{kp} = \frac{P^2}{3}\left(\frac{2}{E_{G}} + \frac{1}{E_{G} + \triangle_{0}}\right)\vec{k}^{2} + eV_{tot}(\vec{r}) - \frac{P^2}{3}\left(\frac{1}{E_{G}} - \frac{1}{(E_{G} + \triangle_{0})}\right)\frac{ie}{\hbar}\vec{\sigma}.(\vec{k}\times \vec{k})\\ + e\frac{P^2}{3}\left(\frac{1}{E_{G}^{2}} - \frac{1}{(E_{G} + \triangle_{0})^{2}}\right)\vec{\sigma}.(\vec{k}\times \vec{E}_{tot})
\eeq
In our analysis the derivation of $\vec{k}.\vec{p}$ perturbed Hamiltonian of the accelerated system  is carried out by using $\vec{E}_{tot}$ and $V_{tot} .$
The total Hamiltonian for the conduction band electrons including the effect of acceleration is then given by,
\beq H_{tot} = \frac{\hbar^{2}\vec{k}^{2}}{2m^*} + eV_{tot}(\vec{r}) + (1 + \frac{\delta g}{2})\mu_{B}\vec{\sigma} . \vec{B} + e(\lambda + \delta \lambda)\vec{ \sigma} .(\vec{ k}\times \vec{ E}_{tot}) ,\label{Hkp } \eeq where
$\frac{1}{m^*} = \frac{1}{m} + \frac{2P^2}{3\hbar^{2}}\left(\frac{2}{E_{G}} + \frac{1}{E_{G} + \triangle_{0}}\right)$ is the effective mass and
$\vec{ E}_{tot} = -\vec{ \nabla} V_{tot}(\vec{r}) = \vec{E} - \vec{E}_{a},$ is the effective total electric field of the
inertial  system and $ \lambda = \frac{\hbar^{2}}{4m^{2}c^{2}}$ is the spin orbit coupling strength as considered in  vacuum. Furthermore, the perturbation parameters $\delta g$ and $\delta \lambda$ are given by
\begin{eqnarray}\label{lam}
\delta g &=& -\frac{4m}{\hbar^{2}}\frac{P^2}{3}\left(\frac{1}{E_{G}} - \frac{1}{E_{G} + \triangle_{0}}\right)\nonumber\\
\delta \lambda &=& + \frac{P^2}{3}\left(\frac{1}{E_{G}^{2}} - \frac{1}{(E_{G} + \triangle_{0})^{2}}\right)
\end{eqnarray}
Specifically, the parameter $\delta g$ is related to the renormalized Zeeman coupling strength, whereas $ \delta \lambda$ is responsible for the renormalization of spin orbit coupling. Now one can rewrite the Hamiltonian as
\beq H_{tot} = \frac{\hbar^{2}k^{2}}{2m^*} + eV_{tot} + (1 + \frac{\delta g}{2})\mu_{B}\vec{\sigma} . \vec{B} + e\lambda_{eff}\vec{\sigma} .(\vec{ k}\times \vec{ E}_{tot}) ,\label{eff} \eeq
where $\lambda_{eff}=\lambda+\delta\lambda$ is the effective SO coupling.

We shall note in due course that the parameter $\lambda_{eff} = (\lambda + \delta \lambda)$, which comes into play due to the interband mixing on the basis of ${\vec k}.{\vec p}$ perturbation theory, is responsible for the enhancement of the spin current.

\section{Spin Hall current and conductivities for cubic and noncubic crystal}
We are interested in the generation of spin current through the effective SOI and therefore consider only the
relevant part of the Hamiltonian (for the positive energy solution) of spin $ \frac{1}{2}$ electron for zero external magnetic field as
\begin{equation}\label{12344}
H =  \frac{\vec{p}^{2}}{2m^*} + eV_{tot}(\vec{r})
- \lambda_{eff}\frac{e}{\hbar}\vec{\sigma}.(\vec{E}_{tot}\times \vec{p})
\end{equation}

The semiclassical equation of motion of electron can be defined as
\beq \vec{F} = \frac{1}{i\hbar}\left[m^*\vec{\dot{r}},H \right] + m^*\frac{\partial\vec{\dot{r}}}{\partial t},\eeq with
$ \vec{\dot{r}}  = \frac{1}{i\hbar}[\vec{r}, H].$ Thus from (\ref{12344})
\beq \vec{\dot{r}} = \frac{\vec{ p}}{m^*} -
\lambda_{eff}\frac{e}{\hbar}\left(\vec{\sigma}\times \vec{ E}_{tot}\right)\label{m}\eeq
Finally, the force
\begin{equation}\label{lor1}
\vec{F} = m^*\ddot{\vec{r}} = -e\vec{ \nabla}V_{tot}(\vec{ r})
+ \lambda_{eff}\frac{em^{*}}{\hbar}\dot{\vec{ r}}\times \vec{\nabla}\times
(\vec{\sigma}\times \vec{E}_{tot})
\end{equation}
is the spin Lorentz force with an effective magnetic field $\vec{\nabla}\times
(\vec{\sigma}\times \vec{E}_{tot}).$
Explicitly, the vector potential is  given by
\begin{equation}\label{gauge}
\vec{A}(\vec{\sigma}) =   \lambda_{eff}\frac{m^{*}c}{\hbar}(\vec{\sigma}\times \vec{E}_{tot})
\end{equation}
Later we shall discuss about this spin dependent gauge $\vec{A}(\vec{\sigma}) $ which is closely related to the $AC$ phase and show how the $\vec{ k} . \vec{ p }$  perturbation modifies the corresponding $AC$ phase.
The spin dependent effective Lorentz force noted in eqn.(\ref{lor1}) is responsible for the spin transport of the electrons in the system, and hence responsible for the spin Hall effect of this inertial system. It is clear from (\ref{lam}) that the expression in (\ref{lor1}) i.e the Lorentz force is enhanced due to $\vec{ k} . \vec{p }$ perturbation in  comparison to the inertial spin force studied in \cite{cb}.
From the expression of $\dot{\vec{r}}$  in (\ref{m}) we can write the linear velocity in a linearly accelerating frame with $\vec{ k} . \vec{p }$ perturbation as
\beq
\displaystyle {\dot{\vec{r}} = \frac{\vec{p}}{m} + \vec{v}_{\vec{ \sigma},~{\vec{a}}}} \eeq where
\begin{equation}\label{mnn}
\displaystyle \vec{v}_{\vec{\sigma},~\vec{a}}=
- \lambda_{eff}\frac{e}{\hbar}(\vec{\sigma}\times\vec{ E}_{tot})
\end{equation}
is the spin dependent anomalous velocity term. The anomalous velocity term is related to the spin  current
as $ j^{i}_{s} = e~n~Tr\sigma_{i}\vec{v}_{\vec{\sigma},~\vec{a}}.$ One should note that the velocity depends on $\delta \lambda$ i.e on the spin orbit gap and the band gap energy of the crystal considered. The expression shows for a non zero spin orbit gap, the spin dependent velocity changes with the energy gap. For vanishing  spin-orbit gap, there is no extra contribution to the anomalous velocity for the $\vec{ k} . \vec{ p }$ perturbation.
The spin current and spin Hall conductivity in an accelerated frame of a semiconductor can now be derived by taking resort to the method of averaging
\cite{n,cb}. We proceed  with equation(\ref{lor1}) as
\begin{eqnarray}\label{lor}
\vec{F} = \vec{F}_{0}+ \vec{F}_{\vec{\sigma}}
\end{eqnarray}
where $ \vec{F}_{0}$ and $\vec{F}_{\vec{\sigma}}$ are respectively the  spin independent and the spin dependent parts of the total spin force.
With the help of eqn. (\ref{gauge}) the Hamiltonian (\ref{12344}) can be written as
\beq H = \frac{1}{2m^{*}}(\vec{p} -\frac{e}{c}\vec{A}(\vec{\sigma}))^{2} + eV_{tot}(\vec{r}) \eeq
where $ V_{tot}(\vec{r}) = V(\vec{r}) - V_{a}(\vec{r})$  and  $V(\vec{r})$  is the sum
of the external electric potential $ V_{0}(\vec{r})$ and the lattice electric
potential $ V_{l}(\vec{r}) $. In this calculation we have neglected the terms of $O(\vec{A}^{2}(\vec{\sigma})).$
Breaking into different parts, the solution of
equation (\ref{lor1}) can be written as $\dot{ \vec{r}} = \dot{ \vec{r}}_{0} +
\dot{ \vec{r}}_{\vec{\sigma}}$\cite{n}. If the relaxation time $\tau$ is independent of $\vec{\sigma}$ and  for the constant total electric field  $\vec{E}_{tot}$, following \cite{n,cb} we can write,
\beq \langle\dot{ \vec{r}}_{0}\rangle = -\frac{\tau}{m^*}\left\langle\frac{\partial V_{tot}}{\partial r}\right\rangle =
\frac{e\tau}{m^*}\vec{E}_{eff}, \label{r0dot} \eeq
and
\begin{equation}
~~~~~~~~~~~~~~~~~\left\langle\dot{\vec{r}}(\vec{\sigma})\right\rangle = - \lambda_{eff} \frac{e^{2}\tau^2}{m^*\hbar}
\vec{E}_{eff}\times\left\langle\frac{\partial }{\partial r} \times (\vec{\sigma}
\times\frac{\partial V_{l}}{\partial r} )\right\rangle $$$$ ~~~~~~~~~~~~~~~~~~~~~~~+ \lambda_{eff} \frac{e^{2}\tau^2}{m^*\hbar}\vec{E}_{eff}\times\left\langle\frac{\partial }{\partial r} \times
(\vec{\sigma} \times \frac{\partial V_{\vec{a}}}{\partial r})\right\rangle.\label{30}
\end{equation}
where  $\vec{E}_{eff} = -e \vec{\nabla}\left(V_{0}(\vec{r}) - V_{\vec{a}}(\vec{r})\right).$
We can now derive the spin current by the evaluation of the averages in equation (21) with different symmetry.
\vspace*{.5cm}

\textbf{\underline{Semiconductors with cubic symmetry }}\\

For the case of  semiconductors with cubic symmetry and  constant acceleration  \cite{cb}
\beq \left\langle\dot{\vec{r}}(\vec{\sigma})\right\rangle = \frac{2e^{2}\tau^2\mu}{m^*\hbar}\lambda_{eff}(\vec{\sigma}\times \vec{E}_{eff})
\label{r1dot}\eeq
as in this case the only non zero contribution permitted by symmetry \cite{n, cb} is
\beq \left\langle\frac{\partial^{2}V_{l}}{\partial r_{i}\partial r_{j}}\right\rangle = \mu \delta_{ij},\label{sym}\eeq
with $\mu$ being a system dependent constant.
The total spin current of this inertial system with $\vec{k}.\vec{p}$ perturbation can now be obtained as
\beq \vec{j}_{kp} = e\left\langle\rho^{s}\vec{\dot{r}}\right\rangle = \vec{j}^{o ,\vec{a}}_{kp} + \vec{j}_{kp}^{s,\vec{a}}(\vec{\sigma}) \eeq
The charge component of this current in our accelerated system is  \beq \vec{j}^{o , \vec{a}}_{kp} = \frac{e^{2}\tau \rho}{m^*}(\vec{E}_{0} - \vec{E}_{\vec{a}}) \label{jo} .\eeq
Let us introduce the density  matrix for the charge carriers as
\begin{equation*}\label{dr}
\rho~^{s} = \frac{1}{2}\rho(1 + \vec{n}.\vec{\sigma}),
\end{equation*}
where $\rho$ is the total charge concentration and $\vec{n}=\langle \vec{\sigma}\rangle$ is the spin polarization vector.\\
  Within the $\vec{k}.\vec{p}$ framework, due to the interband mixing the spin current of this inertial system is given by
\begin{equation}\label{32}
\vec{j}_{kp}^{s, \vec{a}}(\vec{\sigma})  = \left(\frac{2e^{3}\tau^{2}\rho\mu}{m^*\hbar}\right)
\left[\frac{\hbar^{2}}{4(m)^{2}c^{2}} + \frac{P^{2}}{3}\left(\frac{1}{E_{G}^{2}} - \frac{1}{(E_{G} + \triangle_{0})^{2}}\right) \right] \left(\vec{n}\times (\vec{E}_{0} - \vec{E}_{\vec{a}}\right)
\end{equation}
or
\begin{eqnarray}\label{kol}
\vec{j}_{kp}^{s, \vec{a}}(\vec{\sigma}) & =& \frac{m}{m^*}
\left[1 + \frac{4m^{2}c^{2}P^{2}}{3\hbar^{2}}\left(\frac{1}{E_{G}^{2}} - \frac{1}{(E_{G} + \triangle_{0})^{2}}\right)\right]
\vec{j}^{s, \vec{a}}(\vec{\sigma})\\
&=&  \frac{m}{m^*}(1 + \frac{\delta \lambda}{\lambda})\vec{j}^{s, \vec{a}}(\vec{\sigma})
\end{eqnarray}
where $\vec{j}^{s, \vec{a}}(\vec{\sigma}) = \frac{\hbar e^{3}\tau^{2}\rho\mu}{2m^3c^2}\left(\vec{n}\times (\vec{E}_{0} - \vec{E}_{\vec{a}}\right)$ is the spin current in an  accelerating frame\cite{cb} without $\vec{k} . \vec{p}$ perturbation.
With $\vec {E}_0=(0,0,E_z\hat{z})$ and ${\vec a} =(0,0,a_z\hat{z})$ we can explicitly derive the spin currents in the $x$ and $y$ directions.
The ratio of spin current in an accelerating system with and without $\vec{k}.\vec{p}$ perturbation is given by
\beq \frac{|\vec{j}^{s, \vec{a}}(\vec{\sigma})_{kp}|}{|\vec{j}^{s, \vec{a}}(\vec{\sigma})|} = \frac{m}{m^*}(1 + \frac{\delta \lambda}{\lambda}).\eeq
The coupling constant $\delta \lambda$ has different values for different materials and  $\lambda$, the coupling parameter in the vacuum  has a  constant value $3.7\times 10^{-6}{\AA}^{2}.$
We tabulate the ratio of spin currents for different semiconductors with cubic symmetry as \\

\begin{tabular}{|*{2}{c|}l|l|l|}
\hline
$E_{G}(eV)$ & $\triangle_{0}(eV)$ & $P(eV\AA)$ & $\delta \lambda (\AA^{2})$ & $\frac{|\vec{j}^{s, \vec{a}}(\vec{\sigma})_{kp}|}{|\vec{j}^{s, \vec{a}}(\vec{\sigma})|}$ \\
\hline
GaAs = 1.519 & 0.341 & 10.493 & 5.3&$   2.154\times10^{7} $ \\ \hline
AlAs = 3.13  & 0.300 & 8.97  &0.318 & $  5.748\times10^{5} $\\ \hline
InSb = 0.237 & 0.810 & 9.641 &523.33 &$   1.0175\times10^{10} $\\\hline
InAs = 0.418 & 0.380 & 9.197 &120 &$  1.41\times10^{9} $ \\
\hline
\end{tabular}\\
\vspace*{.5cm}\\
The table reveals that  in a linearly accelerating frame, how $\vec{k} . \vec{p}$ method is useful for the generation of
large spin current in semiconductors.
In figure $1$ we have plotted the variation of spin current($x$ direction) with acceleration($z$ direction) using equ. (\ref{spin}) and the
values of the Kane model parameters in the table, for three different semiconductors.

Now if we switch off the external electric field, we have the following expression of spin current in our system
\beq \vec{j}_{kp}^{s, \vec{a}}(\vec{\sigma})  = - \sigma^{s,a}_{H, kp}
\left(\vec{n}\times  \vec{E}_{\vec{a}}\right), \eeq
where $\sigma^{s,a}_{H, kp} = \frac{2e^{3}\tau^{2}\rho\mu}{m^*\hbar}\lambda_{eff}$ is the spin Hall conductivity.
If we now consider the acceleration along $z$ direction, the spin current in the $x$ direction becomes \beq  |\vec{j}_{x,kp}^{s, \vec{a}}(\vec{\sigma})|  = \sigma^{s,a}_{H, kp}(n_{y}E_{a,z})\label{spin}\eeq

\begin{figure}
\includegraphics[width=6.0 cm]{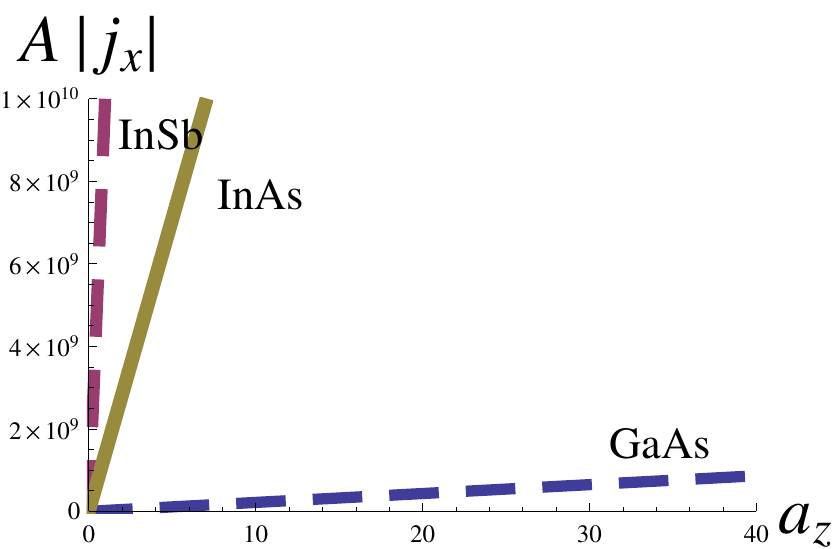}
\caption{\label{a} (Color online) Variation of spin current with acceleration for three different semiconductors, where A = $\frac{2m^{2}c^{2}}{\hbar e^{2}\tau^{2}\rho\mu}$.}
\end{figure}
One can notice that even if the external electric field is zero, still we can achieve huge spin current by the application of acceleration only. For different semiconductors we get different spin
current for non zero spin orbit gap.
It is interesting to point out that for a critical value $\vec{E}_{0} = \vec{E}_{a}$ we see no spin current in the system.
Though the acceleration under which we get the result is very high \cite{cb}, still it elicits that we can control spin current by adjusting acceleration.

The corresponding charge and spin Hall conductivities in a linearly  accelerating frame from the expressions of
the currents (\ref{jo}), (\ref{32}) can be readily obtained as
\begin{eqnarray}\label{mdm}
\sigma^{\vec{a}}_{H, kp} &=& \frac{e^{2}\tau \rho}{m^*}\nonumber\\
\sigma^{s, \vec{a}}_{H, kp} &=& \frac{2e^{3}\tau^{2}\rho\mu}{m^*\hbar}\lambda_{eff}
\end{eqnarray}
As expected \cite{cb}, both the charge and spin conductivities are not affected by the inertial effect of acceleration but the spin conductivity is renormalized by the $\vec{k} .\vec{p}$ perturbation.
The ratio of spin and charge Hall conductivity is
\begin{eqnarray}\label{ratio}
\frac{\sigma^{s , \vec{a}}_{H}, kp}{\sigma^{\vec{a}}_{H, kp}} &=& \frac{2 e\tau\mu}{\hbar}\lambda_{eff}\\
&=& \frac{2 e\tau\mu}{\hbar} \left[\frac{\hbar^{2}}{4(m)^{2}c^{2}} + \frac{P^{2}}{3}\left(\frac{1}{E_{G}^{2}} - \frac{1}{(E_{G} + \triangle_{0})^{2}}\right) \right]
\end{eqnarray}
This spin to charge ratio is independent of the concentration of charge carriers but depends on the relaxation time, the constant term $\mu$, and also on the Kane model parameters $P$, $E_{G}$ and $\triangle_{0}$, i.e. the ratio varies with the material considered.
For a non accelerating system with zero spin orbit gap parameter, the ratio in (\ref{ratio}) is the same as found in \cite{n}.
The condition under which the spin conductivity becomes exactly equal to charge conductivity is
$ \lambda_{eff} = \frac{\hbar}{2 e\tau\mu}. $

The comparison of the spin Hall conductivity in our system with that as obtained in \cite{cb} without $\vec {k}.\vec{p}$ perturbation  shows an enhancement due to the presence of the term $\delta \lambda$ which can be observed from the relation
\beq \frac{|\sigma^{s , \vec{a}}_{H},kp|}{|\sigma^{s, \vec{a}}_{H}|} = \frac{m}{m^{*}}(1 + \frac{\delta \lambda}{\lambda})\eeq
in any crystalline solid with a non zero spin orbit gap $\triangle_{0} .$

\vspace*{.5cm}
\textbf{\underline{Semiconductors with non-cubic symmetry}}\\

There are semiconductors which do not have cubic symmetry, but are examples of producing spin Hall effect. Our objective now is to study those systems and derive the expressions for the spin current. We can consider orthorhombic crystals and can choose the axes of the coordinate frame along the crystal axes \cite{noncubic}.
Instead of eqn (23), for the non cubic symmetry we can now write
\beq \left\langle\frac{\partial^{2}V}{\partial r_{i}\partial r_{j}}\right\rangle = \mu  \chi_{i}\delta_{ij}  ,\label{sym}\eeq
where $\chi_{x} \neq \chi_{y} \neq \chi_{z}$ are the factors of order unity.
Using this,
eqn (\ref{r1dot}) becomes,
\beq \left\langle\dot{\vec{r}}(\vec{\sigma})_{i}\right\rangle = \lambda_{eff}\frac{e^{2}\tau^2}{m^*\hbar} \mu[\chi_{x} + \chi_{y} + \chi_{z} - \chi_{i} ](\vec{\sigma}\times  \vec{E}_{eff})_{i}\label{r2dot}\eeq
which shows that the spin dependent velocity is not uniform in all directions.
Following the same procedure \cite{n}, the spin current in  $ \vec{x} $ direction is attained as
\begin{equation}\label{js}
\vec{j}^{s, \vec{a}}_{x, kp}(\vec{\sigma})  = \lambda_{eff}\left(\frac{e^{3}\tau^{2}\rho\mu}{m^*\hbar}\right) (\chi_{y} + \chi_{z})\left(\vec{n}\times (\vec{E}_{0} - \vec{E}_{\vec{a}})\right)_{x}
\end{equation}
Hence, the spin Hall  conductivity is
\begin{equation}\label{cons}
\sigma^{s, \vec{a}}_{x,kp}  = \lambda_{eff}\left(\frac{e^{3}\tau^{2}\rho\mu}{m^*\hbar}\right)(\chi_{y} + \chi_{z})
\end{equation}

The charge conductivity remains the same as in the cubic case i.e
\beq \sigma^{\vec{a}}_{H, kp} = \frac{e^{2}\tau \rho}{m^*}  \eeq
For an orthorhombic crystal in the $ \vec{x} $ direction we can find out the ratio of the spin to charge conductivity as
\begin{eqnarray}\label{ratio1}
\frac{\sigma^{s , \vec{a}}_{H, kp}}{\sigma^{\vec{a}}_{H, kp}} &=& \lambda_{eff}\frac{ e\tau\mu(\chi_{y} + \chi_{z})}{\hbar}\\
&=& \frac{e\tau\mu(\chi_{y} + \chi_{z})}{\hbar}\\
&&\left[\frac{\hbar^{2}}{4m^{2}c^{2}} + \frac{P^{2}}{3}
\left(\frac{1}{E_{G}^{2}} - \frac{1}{(E_{G} + \triangle_{0})^{2}}\right) \right].
\end{eqnarray}
The charge to spin conductivity ratio does not depend upon the concentration of charge carriers, rather it depends on the Kane model parameters
and the values of $\chi_{x}$ and $\chi_{y}.$
It can be readily observed that we return back to the result of the cubic case for $\chi_{x} = \chi_{y} = \chi_{z}$.

\section{Spin current and spin polarization with time dependent acceleration}
Let us now analyze the case for a time dependent acceleration. As an example of time dependent acceleration we consider \cite{c,cb} \beq \vec{a} = u\omega_{a}^{2}exp(i\omega_{\vec{a}}t)\vec{e}_{x} ,\label{atime}\eeq where the acceleration is induced by harmonic oscillation with frequency $\omega_{\vec{a}}$ and amplitude $u.$ The time dependent acceleration $\vec{a}$ induces a time dependent electric field $\vec{E}_{\vec{a}}$ as  \beq\vec{E}_{\vec{a}} = \frac{mu\omega^{2}_{a}}{e} exp(i\omega_{\vec{a}}t)\vec{e}_{x} \label{eatime}.\eeq
For the external electric field $\vec{E} = 0$, from (\ref{32}), the spin current for a semiconductor with cubic symmetry is then given by
\begin{equation}\label{jtime}
\vec{j}^{s, \vec{a}}_{kp}(\vec{\sigma},t)  = -mu\omega_{a}^{2}\left(\frac{2e^{2}\tau^{2}\rho\mu}{m^*\hbar}\right)
\left[\frac{\hbar^{2}}{4(m)^{2}c^{2}} + \frac{P^{2}}{3}\left(\frac{1}{E_{G}^{2}} - \frac{1}{(E_{G} + \triangle_{0})^{2}}\right) \right]
\left(\vec{n}\times e_{x}\right)exp(i\omega_{\vec{a}}t)
\end{equation}
The $z$ polarized spin current along y direction is
\begin{equation}\label{jtm}
\vec{j}^{s, \vec{a}}_{z, kp}(\vec{\sigma},t)  = -mu\omega_{a}^{2}\left(\frac{2e^{2}\tau^{2}\rho\mu}{m^*\hbar}\right)
\left[\frac{\hbar^{2}}{4(m)^{2}c^{2}} + \frac{P^{2}}{3}\left(\frac{1}{E_{G}^{2}} - \frac{1}{(E_{G} + \triangle_{0})^{2}}\right) \right]
exp(i\omega_{\vec{a}}t)\vec{e}_{y}
\end{equation}
The absolute value of the current is then given by
$|j^{s, \vec{a}}_{z, kp}(\vec{\sigma})| = \frac{1}{A}u\omega_{a}^{2}(1+ \frac{\delta\lambda}{\lambda}),$ where $A = \frac{2m m^{*}c^{2}}{e^{2}\tau^{2}\rho\mu\hbar}$.
Equation (\ref{jtm}) demonstrates that if the semiconductor sample is attached to a mechanical resonator and vibrated in $x$ direction with $\omega_{a} = 10 GHz$ and $u = 10 nm$, we get the $z$ polarized ac spin current along $y$ direction.  Application of $\vec{k} . \vec{p}$ perturbation enhances the ac spin current in semiconductor \cite{c}. In figure $2$ we plot the variation of spin current with amplitude $u$ and frequency $\omega_{a}$ for the $GaAs$ semiconductor.

\begin{figure}
\includegraphics[width=6.0 cm]{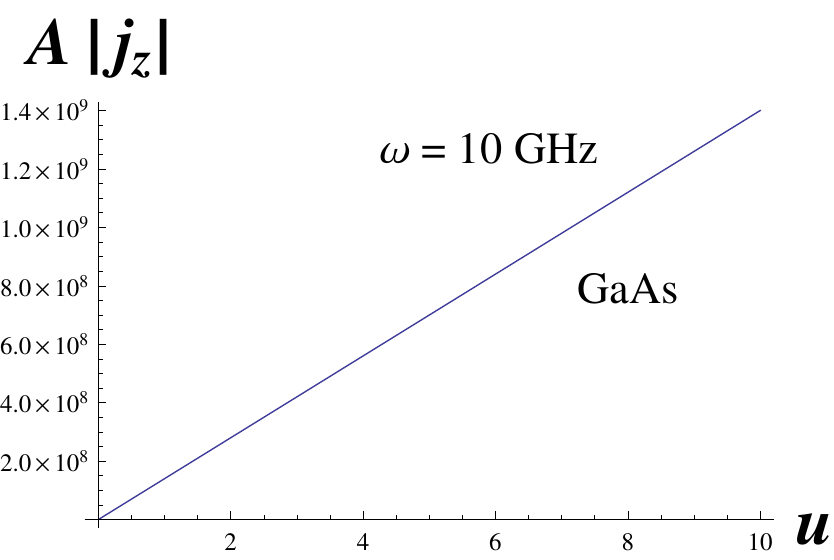}
\hspace*{1cm}
\includegraphics[width=6.0 cm]{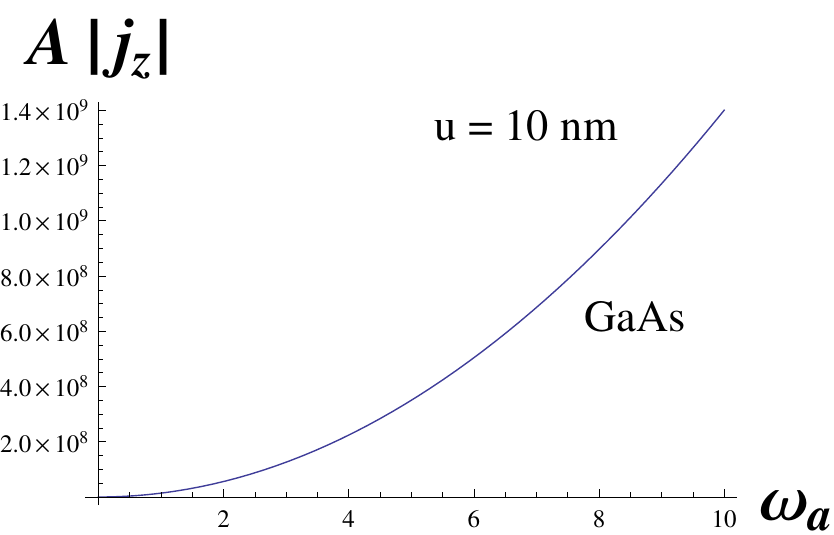}
\caption{\label{b} (Color online) Left: Variation of $A|j_{z}|$ with $u$ for $\omega_{a} = 10 GHz$ . Right: Variation of $A|j_{z}|$ with $\omega_a$ for $u = 10$ nm for $GaAs$ semiconductor, where $A = \frac{2m m^{*}c^{2}}{e^{2}\tau^{2}\rho\mu\hbar}$.}
\end{figure}

Now we move towards the evaluation of the out of plane spin polarization. The constant acceleration of the inertial system cannot explain the out-of-plane transverse spin current
and in what follows we consider a time dependent acceleration 
within the  $\vec{ k} . \vec{ p}$ perturbation formalism. From (11) we write the time dependent Hamiltonian for the time dependent acceleration with the choice of $ \vec{a}(t)= (0, 0, a_{z}\hat{z}(t))$, which subsequently  results $ \vec{ E}_{\vec{ a}}(t) = (0,0, E_{a,z}\hat{z}(t)),$ \cite{cb} and
\beq H(t) = \frac{\hbar^{2}\vec{k}^2}{2m^*} + \alpha_{eff}(k_{x, t}\sigma_{y} - k_{y,t}\sigma_{x}) ,\label{rashba}\eeq
where we use the fact that, for electrons moving through a lattice, the electric field $\vec{E}$  is Lorentz transformed to an effective magnetic field $(\vec{k}\times \vec{E})\approx \vec{B}(\vec{k})$ in the rest frame of the electron.
Hamiltonian (\ref{rashba}) resembles to the well known Rashba Hamiltonian
and $ \alpha_{eff}, $ the spin orbit coupling strength depends on the acceleration of the system as well as on the material parameters.
This Rashba like coupling parameter has  significant importance
in the understanding of the spin transport with inertial effects.
The SOC in semiconductor causes electron to experience an effective momentum dependent magnetic field $\vec{B}_{\vec{a}}(\vec{k}),$ which breaks the spin degeneracy of electron.
Time dependence of the spin orbit Hamiltonian will generate an additional component \cite{fujita} $\vec{ B}_{\perp} = ({\dot\vec{n}}_{\vec{a}}\times \vec{n}_{\vec{a}}),$ in addition to the effective magnetic field $\vec{B}_{a}(\vec{ k}),$
where the unit vector $\vec{n}_{a} = \frac{\vec{B}_{a}(\vec{k})}{|\vec{B}_{a}(\vec{k})|}.$
Let us now assume the effective electric field due to acceleration is in the $x$ direction such that $\vec{ E}_{\vec{ a}} = E_{\vec{a},x}\hat{x}$ \cite{cb,fujita}. As $ \vec{B}_{\vec{a}}(\vec{ k})$ is in the x-y plane, the term $\vec{ B}_{\perp}$ completely represents an effective out of plane magnetic field component along $z$ direction.
With $\vec{ B}_{\Sigma}=\vec{ B}_{\vec{ a}}(\vec{ k})+\vec{ B}_{\perp},$ the total contribution of magnetic field,
the classical spin vector in the $z$ direction can be written as \beq s_{z} = \pm \frac{1}{|\vec{B}_{\Sigma}|}\frac{\hbar}{2}({\dot\vec{n}}_{\vec{ a}}\times \vec{ n}_{\vec{ a}}).\hat{z} \eeq
Now if we make a choice for the unit vector along $\vec{ B}_{a}(\vec{ k})$  as $\vec{n}_{\vec{a}} = p^{-1}(p_{y},-p_{x},0),$  we get $\dot{\vec{ n}}_{a} = p^{-1}(0,e\vec{E}_{\vec{ a}, x}, 0).$
Here $\pm$ represents the spin aligned parallel and anti-parallel to $\vec{ B}_{\Sigma}.$ In the adiabatic limit,
where $\vec{B}_{\vec{a}}(\vec{k})\gg \vec{ B}_{\perp}$ \cite{cb}, $\vec{ B}_{\Sigma}$ approaches $\vec{ B}_{\vec{ a}}(\vec{k})$ and the out of plane spin polarization can be derived as
\begin{eqnarray}\label{szt}
s_{z,kp}\approx &\pm& \frac{1}{|\vec{B}_{\vec{a}}(\vec{k})|}\frac{\hbar}{2}({\dot\vec{n}}_{\vec{a}}\times \vec{n}_{\vec{a}}).\hat{z} \nonumber\\
&=& \pm\frac{\hbar^{2}}{2 \alpha_{eff} p}\frac{\hbar}{2}\left(-\frac{1}{p^{2}}eE_{a ,x}p_{y}\right) \nonumber\\
&=& \mp \frac{e\hbar^{3}p_{y}E_{a ,x}}{4\alpha_{eff} p^{3}}.
\end{eqnarray}
Substituting the value of $E_{a ,x}$ from (\ref{eatime}) in (\ref{szt}), we obtain
the absolute value of $s_{z,kp}$ as \beq |s_{z,kp}| = \mp \frac{e\hbar^{3}p_{y}u\omega^{2}_{a}}{4 \lambda_{eff}a_{z} p^{3}}\label{saz}.\eeq This shows how the Kane model parameters modify the spin polarization vector in an accelerated system \cite{cb}. The dependence of spin polarization on acceleration and Kane model parameters is clear from (\ref{saz}).

In figure $3$ we show the variation of the spin polarization with respect to acceleration $a_z$ and any one of the parameters, $u,\omega_{a}, \frac{p_y}{p^{3}}$ keeping the two other parameters fixed. Here $A = \frac{4\lambda_{eff}}{e\hbar^{3}}$, depends on the solid considered.
\begin{figure}
\includegraphics[width=5.7 cm]{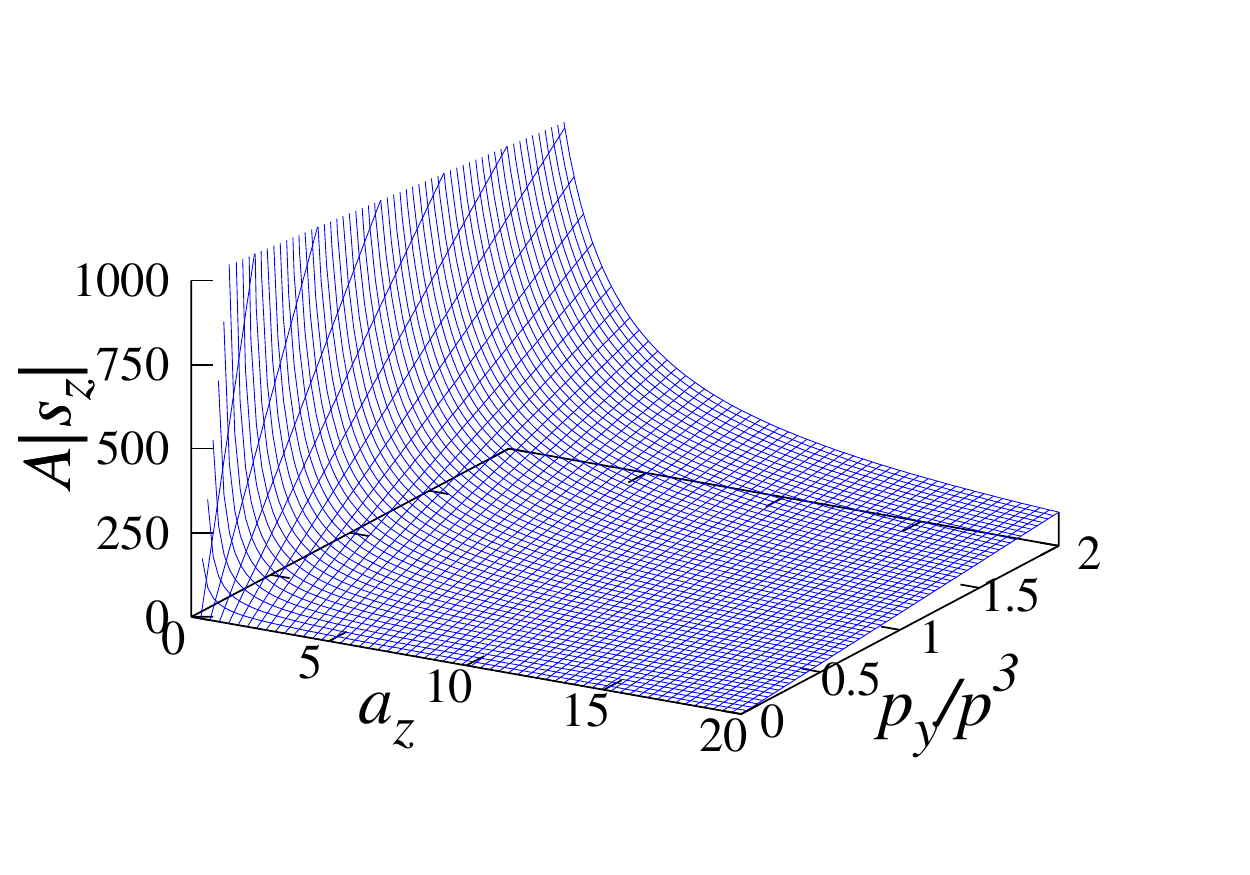}
\includegraphics[width=5.7 cm]{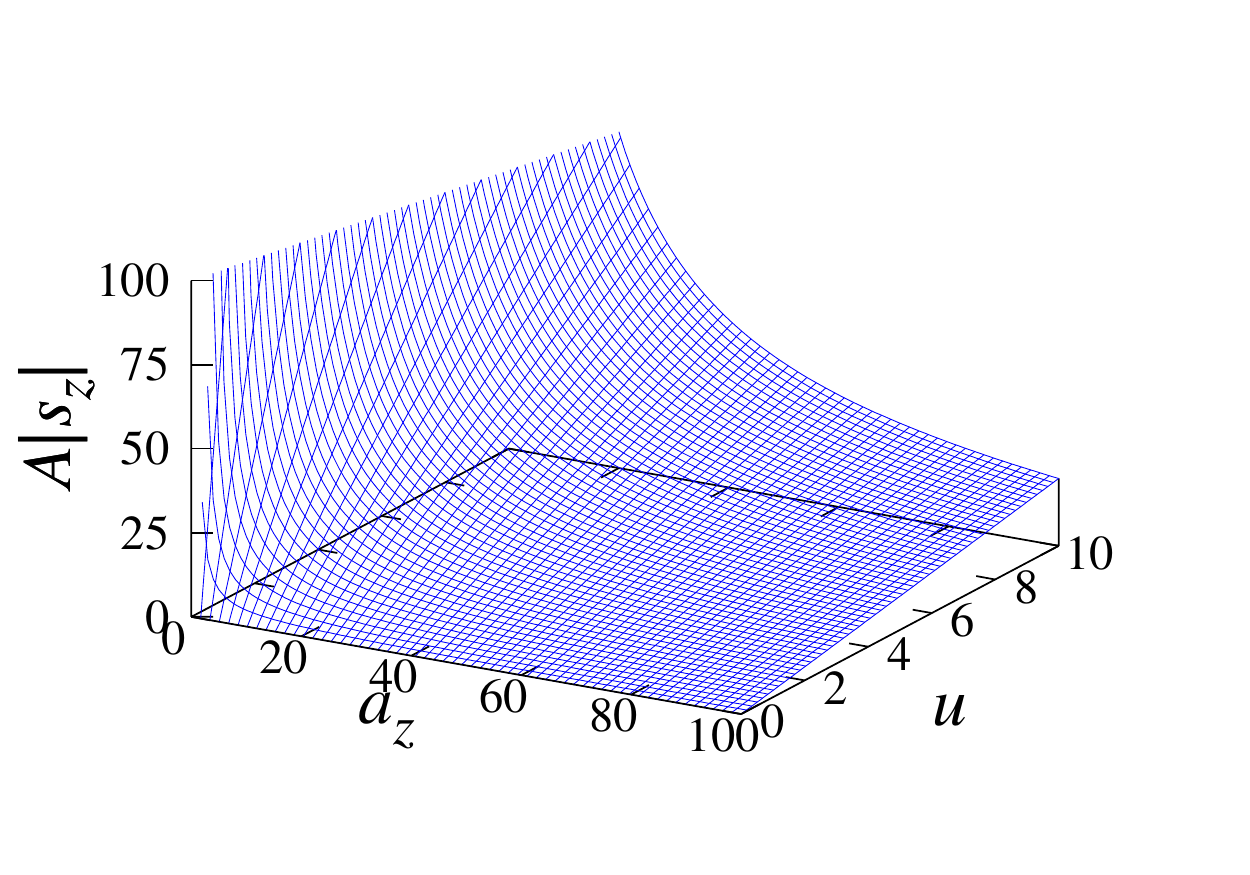}
\hspace{.5cm}
\includegraphics[width=5.7 cm]{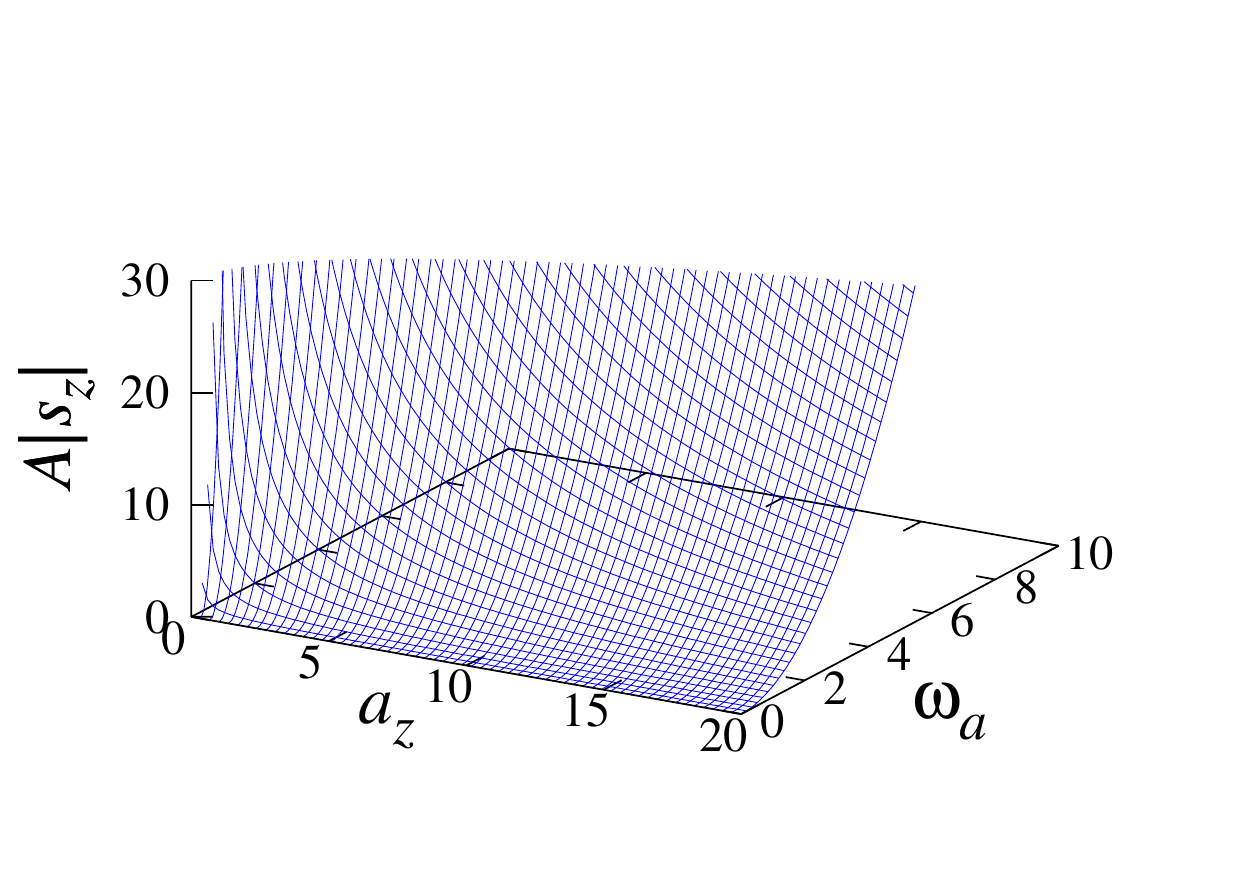}
\caption{\label{c} (Color online) (i) Variation of $A|s_{z}|$ with $a_{z}$ and $p_{y}/p^{3}$ for $\omega_{a}$ = 10 GHz, $u$ = 10 nm.
(ii) Variation of $A|s_{z}|$  with $a_{z}$ and $u$ for $\omega_{a} = 10$ GHz and $\frac{p_{y}}{p^{3}} = const$. (iii) Variation of $A|s_{z}|$  with $a_{z}$ and $\omega_{a}$ for $u = 10$ nm and $\frac{p_{y}}{p^{3}} = const$ where $A = \frac{4\lambda_{eff}}{e\hbar^{3}}$}.
\end{figure}
 As spin polarization is a measurable quantity,
from the experimentally obtained values of $s_{z}$ using (\ref{saz}) we can have an insight for the experimental verification of the parameters of the Kane model.

\section{Gauge field theory of inertial SOC and spin filter}
The study of gauge fields in spintronics has become a topic of recent interest \cite{bliokh}.
In the context of SOI, the importance of Berry phase \cite{berry} was realized following the discovery of the
intrinsic spin Hall effect\cite{shen}. The Berry phase results from cyclic, adiabatic transport of quantum states with respect to
parameter space (e.g. real space ${\vec{r}},$ momentum space ${\vec{k}}).$ In this regard, analysis of the Aharonov-Casher phase through the spin dependent gauge potential also has remarkable importance. Our next goal is to explore the conditions
 of the Berry curvature and study their consequences in spin transport.
In this section we consider the Dirac Hamiltonian in a linearly accelerating frame without any external electric field and consider the physical consequences appearing as a result of the the induced inertial electric field due to acceleration.

\subsection{Spin orbit coupling, spin dependent phase and perfect spin filter}
There are lots of attempt to describe spin filter in different systems \cite{filter2}, but as far as our knowledge goes, proposal of a perfect filter through an inertial system is not noted in the literature.
As the name suggests, the function of a spin filter is to spin polarize the injected charge current.

In this subsection we consider the induced spin orbit Hamiltonian (11) in the presence of the external magnetic field and study the gauge theory of the inertial spin orbit interaction.

In this regard, let us consider the SO Hamiltonian in (\ref{rashba})(time independent) in the presence of external magnetic field $\vec{B}$,
as
\beq  H = \frac{\vec{\Pi}^{2}}{2m^*} + \frac{\alpha_{eff}}{\hbar}(\Pi_{x}\sigma_{y} - \Pi_{y}\sigma_{x}) ,\label{h} \eeq
where $\vec{\Pi} = \vec{p} - e \vec{A}(\vec{r}),$ and $\vec{B} = \nabla\times \vec{A}(\vec{r}). $ The Hamiltonian in (\ref{h}) can also be rewritten in the following form
\beq  H = \frac{1}{2m^{*}}\left(\vec{p} - \frac{e}{c}\vec{A}(\vec{r}) - \frac{q}{c}\vec{A}^{ ~'}(\vec{r}, \vec{\sigma})\right)^{2},\label{min}\eeq
where the spin dependent real space gauge field,  $\vec{A}^{ ~'}(\vec{r}, \sigma)$ is
 \beq \vec{A}^{~ '}(\vec{r}, \sigma) = \frac{c}{2}(-\sigma_{y}, \sigma_{x}, 0) \label{moo}.\eeq Here we neglect the second order of $\vec{A}^{~ '}$ in deriving eqn. (\ref{min}).
The new constant term $q = \frac{2m^*\alpha_{eff}}{\hbar}$ can be regarded as charge and $ \alpha_{eff}$
represents a Rashba \cite{rashba} like spin orbit coupling strength \cite{cb}.
The expression of the spin dependent gauge indicates that it is non-Abelian in nature, whereas the gauge due to external magnetic field provides an Abelian contribution.
The equation (\ref{h}) can be written in terms of the total gauge field $\vec{\tilde{A}}^{~'}(\vec{r}, \sigma)$ 
acting on the system as
\beq  H = \frac{1}{2m^{*}}\left(\vec{p} - \frac{\tilde{e}}{c}\vec{\tilde{A}}\right)^{2} \eeq
 where $\vec{\tilde{A}} = e\vec{A}(\vec{r}) + q\vec{A}(\vec{r}, \vec{\sigma}),$ is the total gauge effective in the system and $\tilde{e}$ is a coupling constant which is set to be $1$ for future convenience.
From the field theoretical point of view, the physical field generated due to the presence of the total gauge $\vec{\tilde{A}}$ is given by
 \beq \Omega_{\lambda} = \Omega_{\mu \nu} = \partial_{\mu}\tilde{A}_{\nu} - \partial_{\nu}\tilde{A}_{\mu} -\frac{i\tilde{e}}{c\hbar}\left[\tilde{A}_{\mu}, \tilde{A}_{\nu}\right] \eeq
The field in the $z$ direction is then
\beq \label{omega} \Omega_{z} =  \left(\partial_{x}\tilde{A}_{y} - \partial_{y}\tilde{A}_{x}\right) - \frac{i\tilde{e}}{c\hbar}\left[\tilde{A}_{x}, \tilde{A}_{y}\right].\eeq
As the commutators of different components of the spin gauge $\vec{A}(\vec{r}, \vec{\sigma})$ exists, $\Omega_{z} $ in our case boils down to the following form
\beq \Omega_{z} = eB_{z} + q^{2}\frac{c}{2\hbar}\sigma_{z} \label{abc}.\eeq
The equn (\ref{abc}) can be expressed in terms of the flux generated through area $S$ as
\beq \Omega_{z} = e\frac{\phi_{B}}{S} + q \frac{\phi_{I}}{S}, \label{abd}\eeq
where $\phi_{B} = SB_{z},$ flux due to external magnetic field and $\phi_{I} = Sq\frac{c}{2\hbar}\sigma_{z}$ is the physical field generated due to inertial spin orbit coupling effect.
The first term on the right hand side(rhs) of (\ref{abd}) is the contribution due to the external magnetic field, which causes the $AB$ phase, whereas the second term on the rhs of (\ref{abd}) is the flux due to the physical field, is actually responsible for a $AC$ like phase. This $AC$ like phase is generated when a spin circulates an electric flux. The second term in the expression of $\Omega_{z}$ in (\ref{abc}), actually represents a magnetic field in $z$ direction, with opposite sign for spin polarized along $+z$ direction or $-z$ direction.
As the spin up and spin down electrons experience equal but opposite vertical magnetic fields, they will subsequently carry equal and opposite $AC$ like phase \cite{casher} which can be obtained from \beq \phi_{AC} = \oint d\vec{r}. \vec{A}(\vec{r}, \sigma),\eeq  whereas the $AB$ phase appears due to the first term in (\ref{abc})  is the same for both up and down electrons.

Interestingly, one can take advantage of this acceleration induced  spin orbit Hamiltonian for the proposition of a perfect spin filter. In a semiconductor the interplay between this $AC$ phase due to the induced spin dependent gauge in presence of acceleration and the $AB$ phase due to an external magnetic field can be used to achieve a spin filter \cite{hatano}. If a spatial circuit can be realized such that the up (down) spin electrons acquire an $AC$ phase of $\pi/2$ (-$\pi/2$) and finite magnetic vector potential in the interior of the circuit makes both the up and down electrons attain an $AB$ phase of $\pi/2$, then the output consists of only spin down electrons and a perfect spin filter is set up. The reversal of the direction of the applied magnetic field may switch the polarity of the filter and the output may consist of only spin down electrons. Thus we can propose theoretically a perfect spin filter without any external electric field. The beauty of our result is that without the application of any external electric field, only through the acceleration of the carriers and external magnetic field we can at least theoretically propose a perfect spin filter for our system.


\subsection{Spatially non uniform SOC and tunable spin filter}
Spin transport through the magnetic barriers is a topic of recent interest, which naturally gives us an idea of spin filter \cite{filter}.
Theoretically, spin filtering through the formation of magnetic barrier was first investigated in a trilayer system constructed using the FM stripes and 2DEG \cite{5} .

In this subsection we consider a trilayer structure, within which a semiconducting (SC) channel is sandwiched between two metallic contacts. The SC channel is assumed to be in an accelerated frame with SO coupling strength $\alpha_{eff}$, while within metallic parts have $\alpha_{eff} = 0$ i.e we are considering a spatial discontinuity of the inertial spin orbit coupling strength. This sharp discontinuity in the SO coupling, in turn gives a highly localized effective magnetic field barriers at the interfaces. 
 The Hamiltonian with spatially non uniform spin orbit coupling can be obtained as
\beq H = \frac{\vec{p}^{2}}{2m^{*}} + \alpha_{eff}(\vec{r})(k_{y}\sigma_{x} - k_{x}\sigma_{y}) \label{hb},\eeq
which can be written in the following form
\beq H = \frac{1}{2m^{*}} \left(\vec{p} - \frac{e}{c}\vec{A}(\vec{r},\sigma)\right)^{2}\eeq
where
 \beq \vec{A}(\vec{r}, \sigma) = \alpha_{eff}(\vec{r})\frac{m^{*} c}{\hbar}(-\sigma_{y}, \sigma_{x}, 0) \label{mn},\eeq
is the spin gauge. The trilayer structure consists of metals at $x = 0$ and $x = L$ and in between there exists semiconducting channel \cite{IEEE}. Let us now consider the spatial profile of $\alpha_{eff}$ to be a step function as
\beq \alpha_{eff}(x) = \alpha_{0}\left[\Theta(x) - \Theta(x - L) \right],\label{non}\eeq where $\Theta(x)$ is the unit step function and $L$ is the length of the semiconductor channel. 
Thus we can write the curvature field  using (\ref{non}) as
\beq \Omega_{z}(\vec{r}) = \frac{m^{*} c}{\hbar}\left(\partial_{x}\alpha_{eff}(x)\sigma_{x} - \partial_{y}\alpha_{eff}(y)\sigma_{y}\right) + \alpha_{eff}^{2}\frac{2(m^{*})^{2}ec}{\hbar^{3}}\sigma^{z} \label{ab}.\eeq
As the spin orbit coupling is non uniform, the first term in (\ref{ab}), which is vanishing for an uniform coupling, exists in our case. Finally we can write the curvature in the $z$ direction as
\beq \Omega_{z}(\vec{r}) = \frac{\alpha_{0}m^{*} c}{\hbar}\left[\delta(x) - \delta(x - L) \right]\sigma^{x} + (\alpha_{eff})^{2}\frac{2(m^{*})^{2}ec}{\hbar^{3}}\sigma^{z} \label{rai},\eeq
where $\alpha_{0}$ and $\delta (x)$ are the inertial spin orbit coupling at the barrier of the sample and the Dirac delta function respectively. The first term on the rhs of eqn. (\ref{rai}) appears as the consequence of the spatial discontinuity of $\alpha_{eff}$ and actually gives narrow spikes of magnetic fields at the interfaces. This term is interesting as it gives a $\delta$ Dirac function centered at the interfaces of the trilayer structure. The second term is a physical field different from the effective magnetic field, generated due to the SO coupling.
This narrow magnetic fields are spin dependent as $\sigma^{x} = \pm 1.$ One should notice here that if  the mixing of spin states are not considered, i.e if we take the length of the channel large compared to the spin precession length, we can write the non- Abelian gauge in (\ref{mn}) as an Abelian gauge as
\beq A_{\pm} = (0, A_{\pm, y}, 0) = (0, \pm \alpha_{eff}\frac{m^{*} c}{\hbar},0 ), \eeq
where $\pm$ denotes two states corresponding to  $\sigma^{x} = \pm 1.$
This structure is useful as a tunable source of spin current, which is very important concept in spintronics applications.

 \section{Conclusion}
In this paper we have theoretically investigated the generation of spin Hall current in a linearly accelerating semiconductor system in presence 
of electromagnetic fields with the help of well known Kane model by taking into account the interband mixing on the basis of $\vec{k} . \vec{p}$ perturbation theory. The explicit form of inertial spin Hall current and conductivity is derived for both cubic and noncubic crystals. We have shown how the interband mixing explains the spin current in an inertial system and also show the dependence of conductivity on the  $\vec{k} . \vec{p}$ perturbation parameters. 
In the case of time dependent acceleration with $\vec{k} . \vec{p}$ method we show the explicit expression of spin current and spin polarization.  From the gauge theoretical point of view, next we have investigated the real space Berry curvature appearing in an inertial system with $\vec{k} . \vec{p}$ method. Lastly based on the gauge theoretical aspects we have discussed a perfect spin filter and a tunable spin filter in an inertial frame.


\end{document}